\documentclass[sigconf]{acmart}
\usepackage{mathptmx}       
\usepackage{helvet}         
\usepackage{courier}        
\usepackage{type1cm}        
\usepackage{graphicx}        
\usepackage{multicol}        
\usepackage[bottom]{footmisc}
\usepackage{subfigure}
\usepackage{amsfonts}
\usepackage{multicol}
\usepackage{multirow}
\usepackage{todonotes}
\usepackage[detect-all]{siunitx}
\usepackage{cleveref}

\newcommand{\gt}{\ensuremath >}

\usepackage{adjustbox}
\usepackage{url}
\newcommand\myurl[2]{\url{#1}}
\usepackage{algorithm}
\usepackage{algorithmicx}
\usepackage[noend]{algpseudocode}

\usepackage{breqn}
\usepackage[mathscr]{euscript}
\usepackage{mathtools}
\usepackage{cancel}

\def\mathperiod{.}

\def\nicefrac#1#2{\leavevmode%
	\raise.5ex\hbox{\small #1}%
	\kern-.1em/\kern-.15em%
	\lower.25ex\hbox{\small #2}}

\newcommand*\Setst[2]%
{\left\{\,#1\vphantom{#2} \;\right|\left. #2 \vphantom{#1}\,\right\}}

\newcommand*\Listst[2]%
{\left[\,#1\vphantom{#2} \;\right|\left. #2 \vphantom{#1}\,\right]}

\AtBeginDocument{%
	\providecommand\BibTeX{{%
			\normalfont B\kern-0.5em{\scshape i\kern-0.25em b}\kern-0.8em\TeX}}}

\acmConference[]{}{}{2021}
\acmBooktitle{}
\acmPrice{}
\acmISBN{}

\title{Approximate Privacy-Preserving Neighbourhood Estimations}
\author{Alvaro Garcia-Recuero}
\email{alvaro.garcia@imdea.org}
\affiliation{%
	\institution{IMDEA NETWORKS}
	\streetaddress{Avd. Del Mar Mediterraneo 22}
	\city{Madrid}
	\state{Leganes}
	\country{Spain}
	\postcode{28918}
}

\begin{document}
	
	\begin{abstract}
		Anonymous social networks present a number of new and challenging problems for existing Social Network Analysis techniques. Traditionally, existing methods for analysing graph structure, such as community detection, required global knowledge of the graph structure. That implies that a centralised entity must be given access to the edge list of each node in the graph. This is impossible for anonymous social networks and other settings where privacy is valued by its participants. In addition, using their graph structure inputs for learning tasks defeats the purpose of anonymity. In this work, we hypothesise that one can re-purpose the use of the HyperANF a.k.a HyperBall algorithm --intended for approximate diameter estimation-- to the task of privacy-preserving community detection for friend recommending systems that learn from an anonymous representation of the social network graph structure with limited privacy impact. This is possible because the core data structure maintained by HyperBall is a HyperLogLog with a counter of the number of reachable neighbours from a given node. Exchanging this data structure in future decentralised learning deployments gives away no information about the neighbours of the node and therefore does preserve the privacy of the graph structure.
	\end{abstract}
	
	\maketitle
	\section{Introduction}
	Performing data analysis over large graphs with billions of edges is an important yet challenging task. Previous work has estimated the average distance in the graph between any two users of large social networking sites such as Facebook, resulting in 4.75 hops on average~\cite{boldi2012four}. Such insights on the structure of a graph are often difficult to investigate for the research community, as it is neither possible to obtain all the neighbourhood sets of each node in the social network due to the terms and conditions for users of the social network API, nor practical to process a complete graph of billions of nodes in terms of performance on a laptop. A practical solution is the emerging use of the graph estimation techniques that approximate the value of graph properties at the node level to reduce computation overhead.
	
	In the literature, estimation algorithms propose to calculate neighbourhood information efficiently by scaling distributed computing to approximate massive graph data mining~\cite{palmer2002anf}. Boldi \& Vigna focus on approximating geometric centrality network metrics at scale using an algorithm based on HyperLogLog (HLL) counters and their HyperANF algorithm~\cite{boldi2011hyperanf}. This allows approximation of geometric centrality metrics at a very high speed and acceptable accuracy for certain use cases that do not require exact values.
	
	\paragraph{Contribution}
	In this work we firstly explore the utility of approximate neighbourhood information of a Decentralised Online Social Network (DOSN) with millions of edges. It is the first study to date with a DOSN dataset using estimation functions. We estimate the neighbourhood function of graph over the Mastodon DOSN data provided by~\cite{zignani2018follow}, namely the average shortest path length, with and without the use of approximated estimations. For the latter, we implement and employ the HLL algorithm using Python to provide the HLL counters that are used in HyperBall. We refer the reader to the HyperBall algorithm in Appendix~\ref{appendix:hb} for a more detailed explanation of the work from~\cite{boldi2013core}.
	Firstly, regarding reproducibility of these results we provide our code in Python, which also implements the original algorithm for HyperBall~\cite{zignani2018follow} and there we apply our neighbourhood function. In all cases, we adjust the error bounds of the estimation technique using HLL~\cite{flajolet2008hyperloglog} counters. The expectation for our benchmark here is to see how a smaller graph, in the order of millions instead of billions, behaves in terms of performance first. Our results show error bounds on the order of 20-30\% in the most pessimistic cases of the network graph traversal (with exception of the level 0 of the traversal where only 1 node exists and thus error is always plausible because of the problem of small cardinalities when using HyperLoLog), while less error occurs otherwise in line with the theoretical foundations in the data approximation technique of HLL. Performance gains in large graphs are evident in Table~\ref{tab:performance}.
	Secondly, our small-world results are according to previous work in the area, in which a random graph (R) should have a lesser average shortest path length than a social network graph (G) with community structure. The size of the community should balance the value of such network metrics among these two kind of graphs when the community size grows so that they become nearly the same~\cite{pattanayak2020lengthening} as we observe in Table~\ref{tab:small-world-performance}. We observe that even to a point in the Mastodon dataset, the R/G proportion is actually higher for the random graph. This shows that in the Mastodon dataset, the federated community structure is vastly growing to a size that actually interconnects communities themselves quite easily to reduce average shortest path length. From that second table of results, we confirm that the Mastodon DOSN shows the highest proportion among all the graphs probed.
	
	\section{Methodology}
	Estimation using sketching techniques such as HLL~\cite{flajolet2008hyperloglog} counters are a tool for cardinality estimation of large multisets. In short they can efficiently count the number of different elements in a stream using a single pass over it. A HLL counter uses a hash function producing $b$ bits that maps each element uniformly to one of $2^b$ possibilities. For any produced hash value, say $h$, we use $\text{Prefix}^t(h)$ to denote the first $t$ bits of h and $\text{Rest}^t(h)$ to denote the remaining bits after the first $t$ bits are removed. Further for any bit-string $x$, let $\text{LZ}(x)$ be the number of leading zeros in $x$. With these definitions in hand, we now present the functions for manipulating HLL counters in Algorithm~\ref{alg:hll}. In addition to functions that add an item to the count and extract the current value of the counters, HLL counters allow us to compute lossless cardinality of unions: the intuition is that between two compatible HLL counters, we can just take the maximum counter to obtain the lossless union of two counters, equivalent to counting the union of the two multi-sets at each of the two counters and thus with no added error. This is however not true for taking the minimum for intersections so we use another technique, namely MinHashes, to approach this challenge and obtain good values as well as shown later here.
	
	\begin{algorithm}
		\caption{HyperLogLog Counter Manipulation}
		\label{alg:hll}
		\begin{algorithmic}[1]
			\State $h:D \rightarrow 2^b$, a hash function from the domain of items
			\State $M$ an array of $m=2^t$ counters each initialised to $-\infty$
			\State $\alpha_m$ is a constant that depends on the number of counters
			
			\Function{AddItem}{$M$:counter, $x$:item}
			\State $i \gets \text{Prefix}^t(h(x))$
			\State $M[i] \gets \text{Max}\{M[i], \text{LZ}(\text{Rest}^t(h(x)))\}$
			\EndFunction
			
			\Function{GetCount}{$M$:counter}
			\State $Z \gets \sum\limits_{i=0}^{m-1}2^{-M[i]}$
			\State $E \gets \alpha_mm^2Z$
			\If{$E \leq \frac{5}{2}m$}\label{corr_begin}
			\State Let V be number of registers equal to 0
			\If{$V \neq 0$}
			\State $E \gets m\text{log}\frac{m}{V}$ \Comment{Small range corrections}
			\EndIf
			\EndIf\label{corr_end}
			\State \textbf{return} $E$
			\EndFunction
			
			\Function{Union}{$M$:counter, $N$:counter}
			\State x:hll\_counter
			\For{i:=0 to m-1}
			\State $x[i] \gets \text{Max}\{M[i], N[i]\}$
			\EndFor
			\State \textbf{Return} x
			\EndFunction
		\end{algorithmic}
	\end{algorithm}
	
	Most crucially for this paper, HLL counters provide tight statistical bounds on the measured cardinality. In the limit $n \rightarrow \infty$, the returned count (say $\hat{n}$) is an almost unbiased estimator~\cite{flajolet2008hyperloglog} of the true count (say $n$) with a relative standard deviation $\frac{\sigma}{n} \leq \frac{1.06}{\sqrt{m}}$. Increasing the number of counters therefore decreases the uncertainty in the measurement at the expense of space. Extending the example above, using 8 bits per counter and 128 counters means we can count items from a universe of $2^{256}$ distinct items with a relative standard deviation under $9.3\%$. The low space overheads and tight statistical bounds are the key features of HLL counters that we wish to exploit in this paper. HLL counters however have one flaw, the bounds on their accuracy only apply at high enough counts. One way to tackle this is to introduce small range correction~\cite{flajolet2008hyperloglog}: if any of the counters are found to be zero then the algorithm returns $m\text{log}(\frac{m}{V})$, where $V$ is the number of zero registers (lines \ref{corr_begin}--\ref{corr_end} of Algorithm~\ref{alg:hll}). Existing work on making HLL counters sound for small counts~\cite{hll_engineer} have recommended using a bias correction instead, and as in their appendix work we implement their recommended bias correction algorithm instead of linear counting at small ranges. We use the same bias correction at low counts for all datasets here, see equation~\ref{eq:precision} below.
	
	\begin{equation}\label{eq:precision}
		\overbrace{101\ldots 011}^{p\ \rm{bits}}\overbrace{001\ldots 110}^{64
			- p\ \rm{bits}}
		\mathperiod
	\end{equation}
	
	\section{Results for Neighbourhood Estimation function}
	The Mastodon DOSN dataset contains about 6.5 million direct edge relations and more than 566520 nodes. In our HyperBall implementation~\cite{boldi2013core} we choose the precision of 4 p-bits for HLL. The traversal of each dataset graph uses a depth as parameter to run with no less than 5 as to resemble the degrees of separation in a real social network graph (e.g., Facebook). We test our HyperBall implementation with two centralised social network datasets (Twitter, Facebook) from~\cite{mcauley2012learning}, and one decentralised social network dataset (Mastodon) for the very first time using a neighbourhood Estimation function.
	
	\paragraph{Performance}
	In Table~\ref{tab:performance} we observe that all of the datasets have naturally worse performance with a brute force approach. The results that use the Hyperball as expected show a considerable speedup in calculation of geometric Hyperballs in the order of minutes, compared to hours when using a brute force Breadth First Search (BFS) approach in a commodity Macbook Pro OS X Mojave with 4 cores CPU, 16GB of memory and 500GB SATA disk installed. In addition, we find a embarrassingly parallel number of tasks in the initialisation and computation of the counters for HyperBall, which provides a further speedup opportunity using the \emph{joblib} library as backend using threading with four embarrassingly parallel initialization tasks of the algorithm. This cuts the computation time of the Mastodon graph's HyperBall by half. Overall the theoretical probabilistic guarantees of the HyperBall seem to reduce times further for larger graphs as Mastodon, when approximations are based on the HLL counters.
	
	\begin{table}
		\centering
		\caption{HyperBall computation times in (hh:mins:secs)}
		\begin{adjustbox}{width=0.9\columnwidth}
			\begin{tabular}{|c|c|c|c|c|c|}
				\cline{4-6}
				\multicolumn{3}{c}{}
				&\multicolumn{1}{|c|}{Bfs}                              &\multicolumn{2}{|c|}{HyperBall} \\ \hline
				& \multicolumn{1}{c|}{\textbf{Nodes}}  &
				\multicolumn{1}{c|}{\textbf{Edges}}  & 
				\multicolumn{1}{c|}{\textbf{Sequential}}  &
				\multicolumn{1}{c|}{\textbf{Sequential}} &
				\multicolumn{1}{c|}{\textbf{Parallel}}
				\\
				\hline
				Twitter & 81306 & 2420766 & \gt 1:00:00.00 & 0:02:54.874483 &  0:02:46.414789\\
				Facebook & 4039 & 88234 & 0:00:56.965906 & 0:00:08.249678 & 0:00:08.533745\\
				Mastodon & 566520 & 6493563 & \gt 1:00:00.00 & 0:11:16.773455 & \textbf{0:05:29.926316} \\
				\hline
			\end{tabular}
		\end{adjustbox}
		\label{tab:performance}
	\end{table}
	
	\paragraph{Network metrics}
	Armed with our implementation of the HyperBall algorithm we also study the \emph{path length probability distribution} and the \emph{average path length} over the same networks datasets. For the latter metric we apply our implementation of the approximated algorithm described in HyperBall. This allows us to compute the so called \emph{small world coefficient} for each of the networks we benchmark.
	
	\begin{algorithm}
		\caption{Average path length with HyperBall}
		\label{alg:ppanf}
		\begin{algorithmic}[1]
			\State nr_paths: number of paths of length per node.
			\State max\_t: maximum distance of HyperBall computations is equal to $b$.
			
			\Function{HyperBall}{$G$:graph, $b$:radius of ball, $p$:hll\_c\_prec}
			\State \textbf{return} $HB$
			\EndFunction
			
			\Function{NumNodesDistFrom}{$v$, $t$}
			\If{t = 0}
			\State \textbf{return} 1
			\ElsIf{t >= get_max_t} 
			\State \textbf{return} 0
			\Else{}
			\State \textbf{return }{balls[v][t].size() - self.balls[v][t - 1].size()}
			\EndIf
			\EndFunction
			
			\Function{average\_path\_length}{$G$}
			\For{v $\in G$}
			\For{$t \in$ 1..{max\_t}}
			\State nr_paths[v] = $HB$.\textproc{NumNodesDistFrom}({$v$, $t$})
			\EndFor
			\EndFor
			\EndFunction
			
		\end{algorithmic}
	\end{algorithm}
	
	\paragraph{Small World}
	\begin{table*}
		\centering
		\caption{Time difference for Algorithm~\ref{alg:ppanf} with NetworkX vs HyperBall computing vs approximating three network metrics.}
		\begin{adjustbox}{width=0.9\textwidth}
			\begin{tabular}{|c|c|c|c|c|c|c|c|}
				\cline{4-8}
				\multicolumn{3}{c}{}
				&\multicolumn{1}{|c}{NetworkX} 
				&\multicolumn{4}{|c|}{HyperBall} \\
				\hline
				& \multicolumn{1}{c|}{\textbf{Nodes}} &
				\multicolumn{1}{c|}{\textbf{Edges}} & 
				\multicolumn{1}{c|}{\textbf{hh:mins:secs}}  &
				\multicolumn{1}{c|}{\textbf{(hh:mins:secs)}} &
				\multicolumn{1}{c|}{\textbf{Avg. Shortest Path Length (R/G)}} &
				\multicolumn{1}{c|}{\textbf{Clustering Coeff.(G/L)}} &
				\multicolumn{1}{c|}{\textbf{Small World Coeff.(l - c)}}
				\\
				\hline
				Twitter & 81306 & 2420766 & 0:03:21.698368 & 0:03:34.580287 & 0.9072928958767731 & 0.565311468612065 & 0.3415490228344126\\
				Facebook & 4039 & 88234 & 0:00:12.997504 & \textbf{0:00:09.321404} & 0.8024902838225895 & 1.0171284634760704 & -0.3015614725029204 \\
				Mastodon & 566520 & 6493563 & 0:15:28.511020 & 0:19:39.377784 & 1.11599944828774786 & $\approx{0.0}$ & 1.1159994828774786 \\
				\hline
			\end{tabular}
		\end{adjustbox}
		\label{tab:small-world-performance}
	\end{table*}
	
	From the computations of Table~\ref{tab:small-world-performance} we obtain the corresponding \emph{small world coefficient} by (i) computing a random graph with equivalent amount of nodes and edges to our input graph, and (ii) computing a lattice graph with same amount of vertices too. Once we have those values computed, now we can obtain the ratio $l$ of average path lengths among a random graph $R$ compared to our actual input graph $G$. If we then also compare the ratio of the average clustering coefficient among the input graph $G$ and the previously computed lattice graph $L$, we obtain the clustering coefficient of $G/L$ called $c$. We calculate the \emph{small world coefficients} as $l - c$. Usually the coefficient should be between -1 and 1 for indicating how strong the small world phenomenon occurs. A coefficient close to 0 indicates a strong influence of small world, negative values indicate regularity/lattice-like graphs, and positive values indicate a graph with more random characteristics. Our result indicates that the Mastodon dataset is of the latter sort of graph according to our results for both raw and approximated values, and due to the random graph effect explained in~\cite{evans2004complex} the clustering coefficient is close or zero indeed for a same random graph of the same number of edges and vertices and even in~\cite{zignani2018follow} it is said that clustering is $\approx{0.17}$ for large in degree nodes and larger for less degree nodes. Also, a fast growing community structure in such a federated social network is plausible and in line with a related preliminary measurements~\cite{complexnets2020}.
	
	We observe that performance of the network properties in the Mastodon dataset improves again somehow. From the standpoint of Social Network Analysis, small-world networks have short average path length and high clustering coefficient (Facebook in Table~\ref{tab:small-world-performance}). On the other hand, an average shortest path length, shorter indicates that information propagates more easily than in a random or regular lattice-type social network as Twitter. For Mastodon this can be related to the relatively self-organising nature of the Mastodon platform among users. For Twitter we would expect a higher average path length as expressed in previous works in the matter~\cite{myers2014information}. However, their dataset at Twitter is more up to data that the one we use from an ego network of 2012 in~\cite{mcauley2012learning}.
		
	Note that the times in Table~\ref{tab:small-world-performance} indicate the average path length computation, which is the main source of overhead for the small world coefficient we provide. Some minor shifts exist over repetition of our runs in Twitter and the Mastodon. We obtain performance times with a parallel initialisation of HLL counters for the HyperBall because it yields a more realistic result than initialising them so sequentially. In the benchmark, for the Mastodon dataset we list in the previous table we compare our approach using HyperANF for getting the path length against state-of-the-art library NetworkX (nx.average\_shortest\_path\_length(G)) that is expected to be more optimised than our prototype. Surprisingly, our approach still performs quite similar or better in some cases for the calculations listed in Table~\ref{alg:ppanf}.
	
	The privacy-preserving version of the average path length above in Algorithm~\ref{alg:ppanf} eventually only needs to learn, as before, the absolute value $|B(v,t)|$ not $B(v,t)$ itself, as $|B(v,t)|$ is the \# of different elements in the summary connected by one edge to the vertex $v$ in question $B(v,t-1)$. Inside HyperBall this is done by first keeping a HLL counter $H(v)$ at each node $v \in V$. So $iff$ $B(v,t) = B(v, t-1)$ holds, then for each $(v,u) \in E, B(v,t) = B(v,t) \cup B(u,t-1)$.
	
	\section{Towards Privacy-Preserving Community Detection}
	Equipped with theoretical vision and empirical results now, we envision a future privacy preserving version of HyperBall for decentralised social network deployments in which network properties require to be exchanged using these set cardinalities or summary sets over insecure channels (encryption suffices) to perform federated learning tasks for instance.
	
	A key operation in community detection is inferring strong ties inside communities by calculating the representative neighbourhood among nodes in a large graph to show the community they form. To provide a timely representation of a community and useful input to other applications as recommendation systems, this can either be done with exact state-of-the-art methods as Louvain~\cite{Louvain} that compare the density of edges in and out modularity, the Hyperball for community detection in~\cite{TriangleCountHyperBall} that estimates the size of those neighbourhoods, counts triangles (note our algorithm differs from theirs also for triangles not shown here). However, we additionally plan to intersect the HyperBalls from several communities whereas one can also potentially decide where to set the ``cut'' based~\cite{schaeffer2007graph} conductance, which yields a ratio of the number of edges connected to nodes outside the cluster in relation to the ones connected to inside nodes. Thus, measuring how well or strong is a community structure. Note this does not consider nor mention anything about privacy of the methods.
	
	However, the HyperBall has a shortcoming as the underlying HLL that it uses can not compute intersections by itself on the fly with the neighbouring summaries or fingerprints alone. Fortunately we can use state-of-the-art estimation techniques based on~\cite{broder1998min} for intersecting two HyperBall. Our approach obtains that by computing the product of the \emph{HLL} with an approximation of the Jaccard similarity using MinHashes. Because MinHashes is approximately the intersection of the Jaccard of the two HyperBall divided by a fixed parameter $k$ (see equation~\ref{eq:minhash}), we are left with just an approximated value of the intersection among two sets in equation~\ref{eq:hll}.
	
	Rather than dividing by the union which would be more costly, for using the MinHashes in the first equation, the result can be approximated as the Jaccard coefficient, which effectively requires just the intersection divided by the MinHashes parameter $k$. Also, this provides desirable privacy properties due to the way the counters and such summary set or fingerprint is built with HLL -- using just an encoding of inputs. It is easy to see from the decomposition below in formulae from equation~\cref{eq:cancelout,eq:minhash,eq:hll} that we can approximately cancel out unions from with the following multiplication of HLL and MinHashes:
	
	\begin{equation}\label{eq:cancelout}
		\frac{|\cap|}{\cancel{|\cup|}} \cdot {\cancel{|\cup|}}
	\end{equation}
	
	\begin{equation}\label{eq:minhash}
		\frac{\left|h_k(A_i) \cap h_k(B_i)\right|}{k}
	\end{equation}
	
	\begin{equation}\label{eq:hll}
		\left|\bigcap A_i \right| = J(A_1,\ldots,A_n)\cdot\left|\bigcup A_i\right| \approx
		\rm{MinHash} \cdot \rm{HLL}
	\end{equation}
	
	Meanwhile, the mentioned MinHashes $H$, if $H'(v) = H(v)$ also holds, then for each  $(v,u) \in E$, $H'(v)$ = $(H'(v) \cap H(u))$. And surprisingly, this can be done while reading edges in sequence.
	
	In summary, the data structure is as follows: (i) 16384, 8 bit entries for the HLL counter. Each entry keeps track of the longest 0 sequence that has been produced for it. From this, we can easily estimate the size of the union of the 2 sets; (ii) 16384, 64 bit hash values in the MinHash table. For each element in the set we generate 16384 hash values with 16384 hash functions and store the smallest ones. Based on how many of these match comparing 2 sets we can estimate the Jaccard coefficient.

	The benefit of our data structure is that it enables fast union with HLL and fast intersection with MinHashes. A disadvantage is that a large dataset may take a long time to generate so many hash values as it is the case in the Mastodon dataset.However, offline pre-computation of the HLL counters is a practical approach.
	
	\section{Conclusion}
	We use approximated computing techniques using summary sets (HyperBall, Minhashes) that can be applied to estimate properties in graphs resulting in algorithms that require only a bounded amount of data, e.g., just a HLL set per vertex rather than a list of all reachable vertices. Summary sets appear to have anonymity properties e.g., hard to deduce actually reachable neighbours given a HLL encoding of neighbours that are reachable in K steps (like in HyperBall) and adding the envisioned signed privacy guarantees in a future work with standard but wisely employed pairing based cryptography~\cite{BLS}. Therefore, moving towards our hypothesis that they HyperBall algorithm is a good candidate for privacy-preserving protocols over decentralised social networks.
	
	Firstly, we may apply our proposal to social recommendation systems or abuse detection in an anonymous social network, where only summary sets are available for friend lists or detection algorithms locally at each node. Therefore, some key features will be the intersection of such lists of friends to make recommendations or detect unsolicited content, both  in zero-knowledge to attaining a secure two party computation by applying existing protocols we have for cardinality estimation and possibly doing so even in a (de)centralised manner.
	
	Secondly, we have explored the feasibility of estimating well-known measures of closeness as the \emph{path length probability distribution} and the \emph{average path length}, the latter using HyperBall for performance. Later we may also consider the variance-to-mean ratio of the shortest-paths distribution in the Mastodon dataset. These metrics are important in network geometry as explained in~\cite{boldi2011hyperanf} and serve as indicator for instance to distinguish structural differences among a social network from a web graph. To the best of our knowledge we are the first to compute them over a lists of ego networks and a decentralised network using approximated or sketching techniques that encode privacy properties.
	
	\paragraph{Limitations}
	A limitation we need to analyse is how the summary sets from the union of HLL counters at each vertex using the candidate set of Algorithm~\ref{alg:ppanf} are vulnerable they are to attacks from an honest but curious participant in the protocols/systems we will develop as said. However if estimated locally these computations are secure, and thus only the broadcasting of the resulting local values to calculate a global aggregated \emph{average path length} need to be protected for privacy and security reasons. Effectively, our network model from here will assume a decentralised computation of such network metrics in future DOSN deployments as Mastodon.
	
	\paragraph{Future work}
	In upcoming work we will show that it is possible to perform efficient community detection with estimation algorithms that use a representation of the neighbouring nodes when computing an intersections/union in a federated manner with privacy. Indeed we would envision the integration of this new work with our former~\cite{garcia2016privacy} in order to intersect fingerprints of the HyperBall in zero-knowledge under standard techniques for exchanging insights of a neighbourhood estimation function across nodes and use it to propagate trust.
	
	\appendix
	\section{Code repository}
	The code of our implementation is available with datasets for reproduceability at github.com/algarecu/ppanf. In order to run our code we use ``pipenv'', a standard modern package manager for python which stores package versions as graph dependencies in a Pipfile and Pipfile.lock files for safe reproduceability of experiments.
	
	\begin{figure}
		\centering
		\includegraphics[scale=0.40]{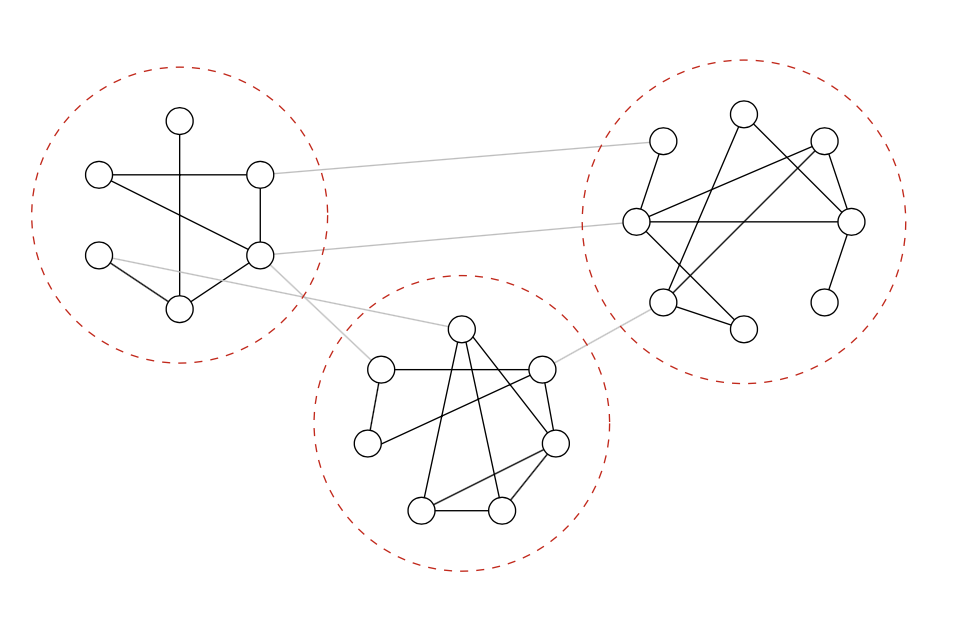}
		\caption{A small-world network community structure of the type considered in the paper from Newman-Girvan~\cite{newman2004finding}.}\label{fig:small-world}.
	\end{figure}
	
	\section{HyperBall algorithm}\label{appendix:hb}
	The HyperBall~\cite{boldi2013core} algorithm uses one HLL per counter per node as required for each iteration in the algorithm here (counter c). In each of those iterations \emph{r}, the size of the counter is approximated represented as a ball absolute value at a vertex, which is essentially $|B_r+1 (v)|$. Note that HyperBall is an adaptation of the HyperANF algorithm and thus our name for the framework we implementing being called ``ppanf'' (privacy preserving ANF). This framework will be useful to several applications we envision as determining approximated triangle count, link prediction and others as even data provenance problems in computer networks with super nodes or hubs acting as neighbourhood estimation vantage points for probing possible paths of trusted information diffusion up/down the network.
	
	\begin{algorithm}
		\caption{Algorithm 3 The HyperBall algorithm as described in ~\cite{boldi2013core}, which returns an estimation of the ball
			cardinality for each node. The functions AddItem and GetCount of Algorithm~\ref{alg:hll} are used.}\label{alg:hyperball}
		
		\begin{algorithmic}[1]
			\State $c[-]$: an array of n HLL counters.
			\Function{Union}{$M$: counter, $N$: counter}
			\For{i < p}
			\State \{$M[i] \gets$ max{$M[i], N[i]$} \}
			\EndFor
			\EndFunction
			\State $r \gets 0$
			\Function{GetBall}{$c$: counter}
			\Repeat
			\For{v $\in$ V}
			$a \gets c[v]$
			\For{w $\in$ N(v)}
			\State $a \gets UNION(c[w], a)$
			\EndFor
			\State write <v, a> to disk, which estimates $|B_r+1 (v)|$.
			\EndFor
			\State Update the array $c[-]$ with the new <v, a> pairs.
			\State$r \gets r+1$
			\Until{no counter changes its value}
			\State \textbf{return }{GetCount(c)}
			\EndFunction
			
			\For{v $\in$ V}\Comment{Initialisation}
			\Function{AddItem}\{$c[v]$, $v$\}
			\EndFunction
			\EndFor
			\State $|\hat{B}_r|_{(r>1)}$ = \Call{GetBall}{$c$: counter}\Comment{Use it}
		\end{algorithmic}
	\end{algorithm}
	
	\section{Related work}
	Small-world networks have a number of properties for which we refer the reader to~\cite{kleinberg2000navigation} and their representation looks often like in those of the state-of-the-art Figure~\ref{fig:small-world} of ~\cite{newman2004finding}. Generally speaking and according to several prominent articles as~\cite{telesford2011ubiquity}, they are "highly clustered, like regular lattices, yet have small characteristic path lengths, like random graphs.", which also explains why is relevant comparing network properties to that of a random or lattice graph, as networks once thought to exhibit small-world properties may not.
	
	\section{Acknowledgements} We thank you Eiko Yoneki and Amitabha Roy from the Cambridge Computer Laboratory for their inputs on HLL counters and MinHashes. We thank you Ajitesh Srivastava for the proofreading.
	
	\bibliographystyle{acm}
	\small{\bibliography{workshop}}
\end{document}